\documentstyle[12pt,axodraw,epsfig]{article}
\def\ltsim{\lower3pt\hbox{$\, \buildrel < \over \sim \, $}}
\def\gtsim{\lower3pt\hbox{$\, \buildrel > \over \sim \, $}}
\def\be{\begin{equation}}
\def\ee{\end{equation}}
\def\ba{\begin{eqnarray}}
\def\ea{\end{eqnarray}}
\def\ga{\mathrel{\raise.3ex\hbox{$>$\kern-.75em\lower1ex\hbox{$\sim$}}}}
\def\la{\mathrel{\raise.3ex\hbox{$<$\kern-.75em\lower1ex\hbox{$\sim$}}}}
\newcommand{\sect}[1]{\section{#1}\setcounter{equation}{0}}

\newcommand{\bi}[1]{\bibitem{#1}}

\begin{document}
\baselineskip=16pt
\begin{titlepage}
\rightline{OUTP-99-68P}
\rightline{UG--FT--110/99}
\rightline{hep-ph/9912552}
\rightline{December 1999}  
\begin{center}

\vspace{0.5cm}

\large {\bf A Three three-brane Universe:\\[2mm]
 New Phenomenology for the New Millennium ?}
\vspace*{5mm}
\normalsize

{\bf Ian I. Kogan$^a$\footnote{i.kogan@physics.ox.ac.uk}, Stavros
Mouslopoulos$^a$\footnote{s.mouslopoulos@physics.ox.ac.uk}, Antonios
Papazoglou$^a$\footnote{a.papazoglou@physics.ox.ac.uk}}\\{\bf Graham
G. Ross$^a$\footnote{g.ross@physics.ox.ac.uk}} and {\bf Jos\'e
Santiago$^b$\footnote{jsantiag@ugr.es}}

\smallskip 
\medskip 
{\it $^a$Theoretical Physics, Department of Physics, Oxford University}\\
{\it 1 Keble Road, Oxford, OX1 3NP,  UK}
\smallskip  

\medskip 
{\it $^b$Department of Theoretical Physics, Granada University}\\
{\it Avenue Fuente Nueva, Granada, 18071, Spain}
\smallskip  

\vskip0.6in \end{center}
 
\centerline{\large\bf Abstract}

We consider an extension of the Randall-Sundrum model with three parallel
3-branes in a 5-dimensional spacetime. This new construction, apart from
providing a
solution to the Planck hierarchy problem, has the advantage that the SM
fields are confined on a positive tension brane. The study of the phenomenology 
of this
model reveals an anomalous first KK state which is generally much
lighter than the remaining tower and also much more strongly coupled
to matter. Bounds on the
parameter space of the model can be placed by comparison
of specific processes with the SM background as well as by the latest
Cavendish experiments. The model suggests a further exotic possibility if one 
drops the requirement of solving the hierarchy problem. In this case gravity may  
result from the exchange of the ordinary graviton plus an ultralight
KK state and modifications of gravity may occur at both small and
extremely large scales.

\vspace*{2mm} 
%\smallskip\newline

\end{titlepage}

\sect{Introduction}

Recently, there has been considerable interest in theories in which the
SM fields are localized on a 3-brane in a higher dimensional spacetime. 
Depending
on the dimensionality and the particular form of the geometry of this
space, the long standing (Planck) hierarchy problem \footnote{i.e. the hierarchy 
between the Planck scale and the electroweak scale. There remains the problem of 
the hierarchy between the weak and gauge unification scales \cite{dum1}.} can 
find  alternative
resolutions. Furthermore, these models make dramatical phenomenological
predictions which can be directly confronted with current and future
accelerator experiments as well as cosmological observations.

Antoniadis, Arkani-Hamed, Dimopoulos and Dvali \cite{D1,D2,ad1} proposed that we
live on a 3-brane in a $3+1+n$ 
space with fully factorizable 
geometry. The higher dimensional Planck scale $M$ is then related to the
4D Planck scale by $M_{\rm Pl}^2=M^{n+2}V_n$ where $V_n$ is the
compactification volume. Taking the size of
the new $n$ dimensions to be sufficiently large and  identifying the
$(4+n)$-Planck scale with the
 electroweak scale, an hierarchy between the electroweak  and
the Planck scale is introduced. Experimental and
astrophysical constraints demand that $n\geq2$, $M\gtsim 30\rm TeV$ and allow 
new dimensions
even of submillimeter size. However, a new hierarchy must now be
explained, namely the ratio of the large compactification radius to the 
electroweak
scale. Even for six extra dimensions this must be greater than $10^5$.

\begin{figure}[ht]
\begin{center}
\mbox{\epsfig{file=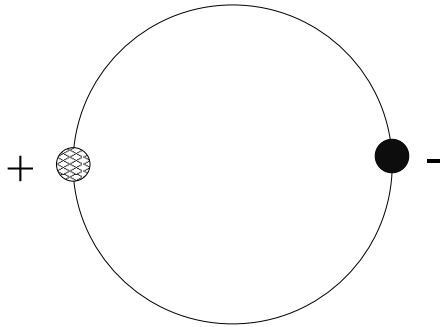,width=6cm}}
\caption{Randall-Sundrum Model with two branes at the orbifold $S^1/Z_2$
fixed points. Here and further a brane with positive cosmological
constant is called positive or  $''+''$ brane and a brane with negative 
cosmological
constant is called  negative or $''-''$ brane.
\label{Fig.1}}
\end{center}
\end{figure}

In the light of this new problem, Randall and Sundrum proposed \cite{RS} an
alternative scenario where they assumed one extra dimension along which
the geometry is
non-factorizable. Their construction consists of two parallel 3-branes sitting
on the fixed points of an $S^1/Z_2$ orbifold (see Fig.~\ref{Fig.1}).
 The 5D spacetime is
essentially a slice of $AdS_5$ and the tensions of the two 3-branes
are chosen so that the 4D spacetime appears flat. This last requirement
forces the one of the two branes to have negative tension. An exponential 
``warp''
factor in the metric then generates a difference of the mass scales between the 
two branes
that could be $\mathcal{O}$$\left(10^{15}\rm GeV\right)$ although the size of 
the
orbifold is only of the order of Planck length. Assuming that the
fundamental mass scale on the positive brane is of the order of
$M_{\rm Pl}$ we can readily get a mass scale on the negative brane of the order
the electroweak scale, thus solving the hierarchy problem. In this the 
compactification radius need only be some 35 times larger than the Planck 
length.
The phenomenology of this model has been extensively explored in
Refs. \cite{Hew,fate}.
 The KK tower of spin-2 graviton resonances starts from the TeV scale with TeV 
spacing giving rise to characteristic signals in high energy
colliders.

Several models have been constructed since then \cite{Oda,manybranes,junc1} that 
extend
the original RS model to multibrane configurations, parallel or
intersecting, with a single or different cosmological constants
between them. A generic characteristic of these models is the presence
of negative tension branes which are necessary for the branes to be
flat. Interesting cases where the branes are not flat or the extra
dimension(s) is not compact have also been considered in
Refs. \cite{Li,Kal,RS2,Lyk2,Diminf}.
 The phenomenology of these models is generally complicated and little has
been written about them. 

For a viable theory the ``brane world'' must reproduce correct
gravity and  cosmology of our Universe.
In  the RS picture the negative cosmological constant of the bulk is 
used to cancel the cosmological constant or tension on the brane. 
On a brane with a positive tension (as for example in  the single brane
scenario \cite{RS2}), gravity is effectively confined to the brane by the
steep ``warp'' factor generated by the tension dominating the brane. 
Of course in realistic cosmologies the energy density 
of the Universe must be dominated by matter. 
An important observation about the cosmology of  the 
``brane world'' was made in Refs. 
\cite{LOW,LK,BDL}. These papers showed that the Hubble
parameter $H$ governing the expansion of the scale factor on the brane
has a different behavior than derived from the usual 4-d Friedmann
equations. In particular, the Hubble parameter is
proportional to  the energy density on the brane instead of the familiar
dependence $H\sim \sqrt \rho$. 

In the papers \cite{NK,Berkl,Cline,Kor}
(see also \cite{moreref}) it was shown that for the Randall-Sundrum construction 
 the full energy density $\rho$ is a sum of a vacuum energy density, 
\textit{i.e.} brane tension
 $\Lambda_{br}$ and a matter energy density $\rho_m$
and the correct expression for Hubble constant squared can be obtained by
cancelling the leading $(\Lambda_{br}+\rho_m)^2$ term  with the term 
$\Lambda_B^2$ coming from the negative bulk 
cosmological constant $\Lambda_B$ so that $H \sim \sqrt{ \Lambda_{br} 
\rho_m}$.
 From this picture one can immediately see that to live on a negative  brane is  
impossible
- we either have normal  matter with positive  energy density, but  imaginary
 Hubble constant, or real Hubble constant, but negative energy density, 
\textit{i.e.} antigravity.
 Because we are living in the Universe with a real Hubble constant and without 
any
 noticeble effects of antigravity the negative  brane as a model of our world  
must be excluded.
It can be shown that one may have normal cosmology on a negative brane
in more general models  with nontrivial bulk stress-energy tensor 
 \cite{kkop,randall3} but  it puts extra constraints on parameters of
the model and in the typical case life on a negative  brane 
leads to the same dilemma - either antigravity or imaginary Hubble constant.

\begin{figure}[ht]
\begin{center}
\mbox{\epsfig{file=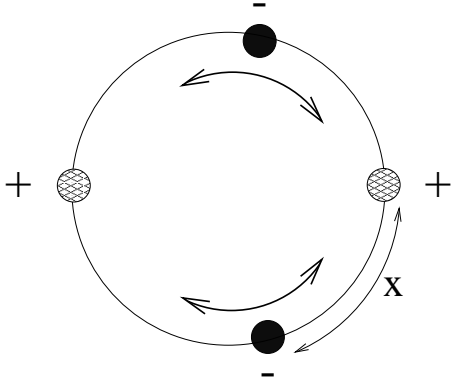,width=6cm}}
\caption { $+-+$ model with two $''+''$ branes at the fixed points and moving
$''-''$ branes. In the limiting case when $x \rightarrow 0$ we have a RS 
configuration.
\label{Fig.2}}
\end{center}
\end{figure}

The purpose of this paper is to formulate a new model of a ``brane world''
 in which we live on a positive brane thus avoiding these cosmological problems. 
The model consists of two positive  branes located at the fixed points of a 
$S_1/Z_2$ orbifold with one negative brane which can move freely in
 between (see 
Fig.~\ref{Fig.2}).
 It is easy to see that the two-brane RS model is nothing but the limiting
 case of our three-brane model, when the negative brane hits one of the 
possitive branes.
 The model has three parameters, the bulk curvature, the warp factor and the $x$ 
factor - which  effectively measures the distance between one of the positive
branes and the negative brane. The  RS model corresponds to the limiting case $x 
= 0$.

\begin{figure}[ht]
\begin{center}
\mbox{\epsfig{file=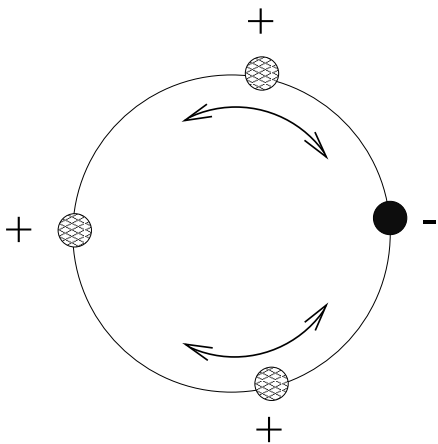,width=6cm}}
\caption{$++-$ model  which is a RS configuration on Figure 1 with an
insertion of  a moving $''+''$ brane. This configuration can be
obtained from a configuration with only $''-''$ branes at the orbifold
points and several moving $''+''$ branes. 
\label{Fig.3}}
\end{center}
\end{figure}

 Let us note that so far it is unclear what are the  selection rules
on branes at orbifold fixed points and moving branes. In a heterotic M-theory
 for example \cite{HetM} one can think about boundaries of
eleven-dimensional space after compactification on a suitable
Calabi-Yau manifold as negative branes where  a fundamental
five-brane in M-theory after wrapping on 2-cycles may play the role of
a $''+''$ brane. Moving a $''+''$ brane towards one of the  $''-''$ branes
 we can transform it into a $''+''$ brane, so one can get a
configuration of the type $++-$  or   with more than one moving
$''+''$ brane - the only constraint is the that the sum of all ``charges'' 
  is zero.  Let us note that each  moving brane on an orbifold must
 be  counted with a double charge because of it mirror image.  
\footnote{We are grateful to Andre Lucas for
interesting discussion on this subject.}
It is unclear if one may have moving $''-''$ branes. At the same time 
it is necessary to say that so far  the full string/M-theory
description of these multibrane configurations is missing and we can
not exclude a possibility that there are moving negative branes.

In this paper we shall discuss in details the $+-+$ model of Fig.2.
One of the striking predictions of our model is the fact that the first KK mode 
can be
 very light and strongly coupled compared to the rest of the KK states, so
that  the phenomenology is determined mostly by it. An unusual
possibility arises if we relax the requirement that the
Planck hierarchy problem is solved. In this case this light mode can be so light 
that the corresponding wavelength
 can be by order of $1\%$ of the observable size of the Universe while
 the {\it second} KK mode is in submillimeter region. Surprisingly enough this 
situation is not excluded experimentally! Thus one may have ``Bi-Gravity'' -  in all
experimentally analyzed regions gravitational attraction is due to an
exchange of two particles - the massless graviton and ultralight  first KK 
mode. Only at scales larger than $10^{26}cm$ will the first KK mode decouple 
leading to a much smaller gravitational coupling beyond this length scale. 

The reason the anomalously light KK mode exists is due to the fact that with 
more that one
$''+''$ brane there will be a bound state on {\it each} of them when they have
infinite separation. At finite distances there is a mixing between the two 
localized states. One superposition  is the true ground
state while the other
configuration has non-zero mass, but the gap may be very small - it
 is given by a tunneling factor.  This effect takes place not only in
 $+-+$ model but in $++-$  and other models  with more than two $''+''$
branes (in which case there may be more than one light mode). In this paper we 
consider only the $+-+$ configuration which has the interesting ``Bi-Gravity`` 
possibility. Other models will
 be considered in separate publications.

The organisation of the paper is as follows. In the next Section we construct 
the model
 and discuss the spectrum of KK excitations. In Section 3 the first
and subsequent KK modes are considered
 in more detail and their masses and couplings as a function of the
``warp'' factor and the parameter
 $x$ are discussed
 using both analytical and numerical methods. In Section 4 we discuss
the phenomenology when the
 first KK mode  has a mass in the meV-TeV region. In Section 5 we 
discuss the unusual ``Bi-Gravity'' scenario
 when the Compton wavelength of the first KK mode is  by order of
$10^{26} {\rm cm}$ while the Compton wavelength of a second KK mode is less than 
$1 {\rm
mm}$. 
  In conclusion, we discuss the 
possible generalization of our model and questions for future investigation.

\sect{The 3-brane model}

Our model consists of three parallel 3-branes in an $AdS_5$ space with
cosmological constant $\Lambda<0$. The 5-th dimension has the geometry
of an orbifold and the branes are located at
$L_0=0$, $L_1$ and $L_2$ where $L_0$ and $L_2$
are the orbifold  fixed
points (see Fig.2). Firstly we consider the branes having no  matter on them in
order to find a suitable vacuum solution. The action of this setup is:
\be
S=\int d^4 x \int_{-L_2}^{L_2} dy \sqrt{-G} 
\{-\Lambda + 2 M^3 R\}-\sum_{i}\int_{y=L_i}d^4xV_i\sqrt{-\hat{G}^{(i)}}
\ee
where $\hat{G}^{(i)}_{\mu\nu}$ is the induced metric on the branes
and $V_i$ their tensions. The notation is the same as in
Ref. \cite{RS}.
 The Einstein equations that arise from this
action are:
\be
R_{MN}-\frac{1}{2}G_{MN}R=-\frac{1}{4M^3}
\left(\Lambda G_{MN}+
\sum_{i}V_i\frac{\sqrt{-\hat{G}^{(i)}}}{\sqrt{-G}}
\hat{G}^{(i)}_{\mu\nu}\delta_M^{\mu}\delta_N^{\nu}\delta(y-L_i)\right)
\ee

At this point we demand that our metric respects 4D Poincar\'{e}
invariance. The metric ansatz with this property is the following:
\be
ds^2=e^{-2\sigma(y)}\eta_{\mu\nu}dx^\mu dx^\nu +dy^2
\ee

Here the ``warp'' function $\sigma(y)$ is essentially a conformal
factor that rescales the 4D component of the metric. A straightforward
calculation gives us the following differential equations for $\sigma(y)$:
\ba
\left(\sigma '\right)^2&=&k^2\\
\sigma ''&=&\sum_{i}\frac{V_i}{12M^3}\delta(y-L_i)
\ea
where $
k=\sqrt{\frac{-\Lambda}{24M^3}}$ is a measure of the curvature of the bulk. 

The solution of these equations consistent with the orbifold geometry is 
precisely:
\be
\sigma(y)=k\left\{L_1-\left||y|-L_1\right|\right\}
\ee
with the requirement that the brane tensions are tuned to $V_0=-\Lambda/k>0$,
$V_1=\Lambda/k<0$, \mbox{$V_2=-\Lambda/k>0$}. 
If we consider massless fluctuations of this vacuum metric  as in
Ref. \cite{RS} and then integrate over the 5-th dimension, we find
the 4D Planck mass is given by
\be
M_{\rm Pl}^2=\frac{M^3}{k}\left[1-2e^{-2kL_1}+e^{-2k(2L_1-L2)}\right]
\ee

The above formula tells us that for large enough $kL_1$ and  $k\left(2L_1-
L_2\right)$ the
three mass scales $M_{\rm Pl}$, $M$, $k$ can be taken to be of the same
order. Thus we take $k\sim {\mathcal {O}}(M)$ in order
not to introduce a new hierarchy, with the additional restriction
$k<M$ so that the bulk curvature is small compared to the 5D Planck
scale so that we can trust our solution. Furthermore, if we put matter on the 
third brane all the physical masses $m$ on the third brane will be related to 
the
mass parameters $m_0$ of the fundamental 5D theory by the conformal (warp) 
factor
\be
m=e^{-\sigma\left(L_2\right)}m_0=e^{-k\left(2L_1-L_2\right)}m_0
\ee

Thus we can assume that the third brane is our universe and get a solution of 
the Planck hierarchy problem arranging
$e^{-k\left(2L_1-L_2\right)}$ to be of
$\mathcal{O}$$\left(10^{-15}\right)$, \textit{i.e} $2L_1-L_2\approx35k^{-1}$. In this case 
all the parameters of the
model $L_1^{-1}$, $L_2^{-1}$ and k are of the order of Plank scale.

To determine the phenomenology of the model we need to know the KK
spectrum that follows from the dimensional reduction. This is
determined by considering the (linear) fluctuations of the metric of the
form:
\be
ds^2=\left[e^{-2\sigma(y)}\eta_{\mu\nu} +\frac{2}{M^{3/2}}h_{\mu\nu}
(x,y)\right]dx^\mu dx^\nu +dy^2
\ee

Here we have ignored the dilaton mode that could be used to stabilize
the brane positions $L_1$ and $L_2$ as discussed in
Refs.\cite{randall3,GW,Wolf,Lut}. While the dilaton and its KK excitations may 
lead to additional observable phenomena its inclusion should not change the 
phenomenology of the graviton KK modes. We have also ignored the off diagonal
vector KK modes simply because they don't couple to the SM fields.

We expand the field $h_{\mu\nu}(x,y)$ in graviton and KK states plane waves:
\be
h_{\mu\nu}(x,y)=\sum_{n=0}^{\infty}h_{\mu\nu}^{(n)}(x)\Psi^{(n)}(y)
\ee
where
$\left(\partial_\kappa\partial^\kappa-m_n^2\right)h_{\mu\nu}^{(n)}=0$
and fix the gauge as
$\partial^{\alpha}h_{\alpha\beta}^{(n)}=h_{\phantom{-}\alpha}^{(n)\alpha}=0$.
The function $\Psi^{(n)}(y)$ will obey a second order differential
equation which after a change of variables reduces to an ordinary
Schr\"{o}dinger equation:
\be
\left\{-
\frac{1}{2}\partial_z^2+V(z)\right\}\hat{\Psi}^{(n)}(z)=\frac{m_n^2}{2}\hat{\Psi
}^{(n)}(z)
\ee

\be
{\rm with}\hspace*{0.5cm} V(z)=\frac{15k^2}{8[g(z)]^2}-
\frac{3k}{2g(z)}\left[\delta(z)+\delta(z-z_2)-\delta(z-z_1)-\delta(z+z_1)\right]
\ee

The new variables and wavefunction in the above equation are defined as:
\be
\renewcommand{\arraystretch}{1.5}
z\equiv\left\{\begin{array}{cl}\frac{2e^{kL_1}-e^{2kL_1-ky}-
1}{k}&y\in[L_1,L_2]\\\frac{e^{ky}-1}{k}&y\in[0,L_1]\\-\frac{e^{-ky}-1}{k}&y\in[-
L_1,0]\\-\frac{2e^{kL_1}-e^{2kL_1+ky}-1}{k}&y\in[-L_2,-L_1]\end{array}\right.
\
\ee
\be
\hat{\Psi}^{(n)}(z)\equiv \Psi^{(n)}(y)e^{\sigma/2}
\ee
and the function $g(z)$ as $
g(z)\equiv k\left\{z_1-\left||z|-z_1\right|\right\}+1$, where $z_1=z(L_1)$.

This is a quantum mechanical problem with $\delta$-function potentials of
different weight and an extra $1/g^2$ smoothing term (due to the AdS geometry) 
that gives the
potential a double ``volcano'' form. The change of variables has been
chosen so that there are no first derivative terms in the
differential equation. 

An interesting characteristic of this potential is that it always
gives rise to a (massless) zero mode which reflects the fact that
Lorentz invariance is preserved in 4D spacetime. It is given by
\be
\hat{\Psi}^{(0)}=\frac{A}{[g(z)]^{3/2}}
\ee

The
normalization factor $A$ is determined by the requirement 
$\displaystyle{\int_{\phantom{.}0}^{z_2}
dz\left[\hat{\Psi}^{(0)}(z)\right]^2=1}$, chosen so that we get the standard 
form 
of the Fierz-Pauli Lagrangian.

In the specific case where $L_1=L_2/2$
(and with zero hierarchy) the potential and thus the zero mode's
wavefunction is
symmetric with respect to the second brane. When the second brane
moves towards the third one the wavefunction has a minimum on the
second brane but different heights on the other two branes, the difference 
generating the hierarchy between the first and the third brane.

For the KK modes the solution is given in terms of Bessel
functions. For $y$ lying in the regions ${\bf A}\equiv\left[0,L_1\right]$ and
${\bf B}\equiv\left[L_1,L_2\right]$, we have:
\be
\hat{\Psi}^{(n)}\left\{\begin{array}{cc}{\bf A}\\{\bf 
B}\end{array}\right\}=\sqrt{\frac{g(z)}{k}}\left[\left\{\begin{array}{cc}A_1\\B_
1\end{array}\right\}J_2\left(\frac{m_n}{k}g(z)\right)+\left\{\begin{array}{cc}A_
2\\B_2\end{array}\right\}Y_2\left(\frac{m_n}{k}g(z)\right)\right]
\ee

The boundary conditions (one for the
continuity of the wavefunction at $z_1$ and three for the
discontinuity of its first derivative at $0$, $z_1$, $z_2$) result in a
$4\times4$ homogeneous linear system which, in order to have a
non-trivial solution, leads to the vanishing determinant:

\be
\renewcommand{\arraystretch}{1.5}
\left|\begin{array}{cccc}J_1\left(\frac{m}{k}\right)&Y_1\left(\frac{m}{k}\right)
&\phantom{-}0&\phantom{-}0\\0&0&\phantom{-
}J_1\left(\frac{m}{k}g(z_2)\right)&\phantom{-
}Y_1\left(\frac{m}{k}g(z_2)\right)\\J_1\left(\frac{m}{k}g(z_1)\right)&Y_1\left(
\frac{m}{k}g(z_1)\right)&\phantom{-}J_1\left(\frac{m}{k}g(z_1)\right)&\phantom{-
}Y_1\left(\frac{m}{k}g(z_1)\right)\\J_2\left(\frac{m}{k}g(z_1)\right)&Y_2\left(
\frac{m}{k}g(z_1)\right)&-J_2\left(\frac{m}{k}g(z_1)\right)&-
Y_2\left(\frac{m}{k}g(z_1)\right)\end{array}\right|=0
\ee
(where we have suppressed the subscript $n$ on the masses $m_n$)

This is essentially the mass quantization condition which gives the
spectrum of the KK states. For each mass we can then determine the wave function 
with normalization $\displaystyle{\int_{\phantom{.}0}^{z_2}
dz\left[\hat{\Psi}^{(n)}(z)\right]^2=1}$. 

From the form of the potential we can
immediately deduce that there is  a second ``bound'' state, the
first KK state. In the symmetric case, $L_1=L_2/2$, this is simply given by 
reversing the sign of the graviton wave function for $y>L_1$ (it has one zero
at $L_1$). When the second brane moves towards the third this symmetry
is lost and the first KK wave function has a very small value on the
first brane, a large value on the third and a zero very close to the
first brane.

The interaction of the KK states to the SM particles is found as in
Ref. \cite{Lyk1} by expanding the minimal
gravitational coupling of the SM Lagrangian $\displaystyle{\int
d^4x\sqrt{-\hat{G}}{\mathcal{L}}\left(\hat{G},SM fields\right)}$ with respect to 
the metric. After the rescaling due to the ``warp'' factor we get:
\ba
{\mathcal{L}}_{int}&=&-\frac{g\left(z_2\right)^{3/2}}{M^{3/2}}\sum_{n\geq
0}
\hat{\Psi}^{(n)}\left(z_2\right)h_{\mu\nu}^{(n)}(x)T_{\mu\nu}\left(x\right)= 
\nonumber
\\&=&-\frac{A}{M^{3/2}}h_{\mu\nu}^{(0)}(x)T_{\mu\nu}\left(x\right)-
\sum_{n>0}\frac{\hat{\Psi}^{(n)}\left(z_2\right)g\left(z_2\right)^{3/2}}{M^{3/2}
}h_{\mu\nu}^{(n)}(x)T_{\mu\nu}\left(x\right)
\ea
with $T_{\mu\nu}$ the energy momentum tensor of the SM
Lagrangian. Thus the coupling suppression of the zero and KK modes to matter is
respectively:
\ba
\frac{1}{c_0}&=&\frac{A}{M^{3/2}}\\
\frac{1}{c_n}&=&\frac{\hat{\Psi}^{(n)}\left(z_2\right)g\left(z_2\right)^{3/2}}{M
^{3/2}}
\ea

For
the zero mode the normalization constant $A$ is
$\frac{M^{3/2}}{M_{\rm Pl}}$ which gives the Newtonian gravitational
coupling suppression $c_0=M_{\rm
Pl}$.

\section{The first  and subsequent KK modes: Masses and coupling constants}

As discussed above the KK spectrum has a special first mode which for all 
$x$ significantly different from unity has very different behaviour
compared to the other KK states. In the case $x>>1$ we may obtain a reliable
approximation to its mass by using the first terms of the Bessel power
series. From now on we  shall use a convenient choice of parameters:
the measure of the curvature of the bulk $k$, the separation of the
second and third brane $x=k(L_2-L_1)$ and the hierarchy factor
$w=e^{-k\left(2L_1-L_2\right)}$. The first KK mode has mass given by 
\be
m_1=2kwe^{-2x}
\label{m1}
\ee

The normalization integral can also then be done
analytically and the coupling suppression is found to be independent of $x$ and 
equal to 
\be
c_1=wM_{\rm Pl}
\label{c1}
\ee

The reason for this is readily understood because, being dominantly a bound 
state of the volcano potential on our brane, it is largely localized on it. 

 The 
masses of the other KK states in the above region are found to
depend in a different way on the parameter $x$. Numerically we find
out that the mass of the
second state and the spacing $\Delta m$ between the subsequent states have the 
form:
\ba
m_2&\approx&4kwe^{-x}\\
\Delta m&\approx&\varepsilon kwe^{-x}
\label{m2}
\ea
where $\varepsilon$ is a number between 1 and 2. The spacing only approaches a 
constant for high enough
levels when the Bessel arguments become much greater than one. Again solving 
numerically, we find the couplings of the higher modes are suppressed relative 
to the lowest mode by a factor proportional to $e^x$. To illustrate the 
difference between the first and second modes at large $x$, we consider the case 
$w=e^{-35}$,
$x=5$ and $k=10^{17}{\rm GeV}$. The masses and coupling suppressions of the 
first 
two modes are
\be
\renewcommand{\arraystretch}{1.5}
\begin{array}{lll}
m_1\approx5.7{\rm MeV}&{\rm with}&c_1=630{\rm GeV}\\
m_2\approx1.6{\rm GeV}&{\rm with}&c_2=52500{\rm GeV}
\end{array}
\ee

When the second brane approaches the
third brane, the above approximations break down and the first mode is
not so different from the others. Its mass rises above $100{\rm GeV}$
and the coupling is no longer constant but has a small dependence on
$x$. In the extreme case where the
positions of the second and the third brane coincide, the positive
brane disappears and we obtain the
original Randall-Sundrum model. In this case the spectrum starts from several
hundreds of GeV up to some TeV depending on the choice of $k$. 
It has large spacing between the KK states and TeV coupling suppression.

\section{Phenomenology}

In this Section we will present a brief discussion of the phenomenology of the 
KK modes to be expected in colliders, concentrating on the simple process 
$e^+e^-\rightarrow\mu^+\mu^-$. The analysis is readily generalized to include 
$q\bar{q}$, $gg$ initial and final states. A more complete discussion will 
appear elsewhere.

\begin{figure}
\begin{center}
\begin{picture}(300,100)(0,50)
\ArrowLine(50,50)(100,100)
\ArrowLine(50,150)(100,100)
\ArrowLine(200,100)(250,150)
\ArrowLine(200,100)(250,50)
\Vertex(100,100){2}
\Vertex(200,100){2}
\LongArrow(140,110)(160,110)
\Photon(100,100)(200,100){4}{6}
\Photon(100,100)(200,100){-4}{6}
\Text(90,125)[]{$e^+$}
\Text(90,75)[]{$e^-$}
\Text(210,125)[]{$\mu^+$}
\Text(210,75)[]{$\mu^-$}
\Text(150,120)[]{$\sqrt{s}$}
\end{picture}
\caption{$e^+e^-\rightarrow \mu^+ \mu^-$}\label{feyn:diag}
\end{center}
\end{figure}
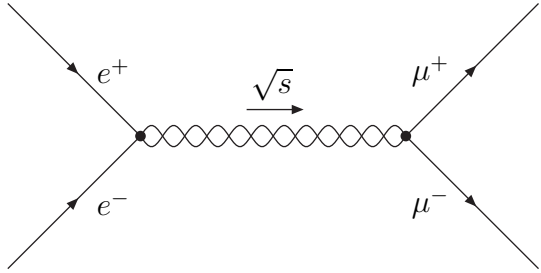 

Using the Feynman rules of Ref.\cite{Lyk1} the contribution of the KK modes to 
$e^+e^-\rightarrow\mu^+\mu^-$ is given by
\be
\sigma\left(e^+e^-\rightarrow\mu^+\mu^-\right)=\frac{s^3}{1280\pi}|D(s)|^2
\ee
where $D(s)$ is the sum over the propagators multiplied by the
appropriate coupling suppressions:
\be
D(s)=\sum_{n>0}\frac{1/c_n^2}{s-m_n^2+i\Gamma_n m_n}
\ee
and $s$ is the C.M. energy of $e^+e^-$.

Note that the bad high energy behaviour of this cross section is expected since 
we are working with an effective - low
energy non-renormalizable theory of gravity. Our effective theory is valid up to 
an energy scale
$M_{s}$ (which is ${\mathcal{O}}({\rm TeV})$), which acts as an ultraviolet 
cutoff. The theory that is valid  above this scale
is supposed to give a consistent description of quantum gravity. Since this is 
unknown we are only able to detemine the contributions of the KK states with 
masses less than this
scale. This means that the summation in the previous formula should
stop at the KK mode with mass near the cutoff.

The decay rates of the KK states that are present in the above formula
are given by:
\be
\Gamma_n=\beta\frac{m_n^3}{c_n^2}
\ee
where $\beta$ is a dimensionless constant that is between
$\frac{9}{80\pi}\approx0.035$ (in the case that the KK is light
enough, \textit{i.e.} smaller than $0.5 {\rm MeV}$, that can decay only to 
massless 
gauge bosons) and $\frac{37}{192\pi}\approx0.061$ (in the case where
the KK is heavy enough that can decay to all SM particles).

The details of computation
of the total  cross section depends on the KK spectrum. In the case that
$x$ is small, $x\ltsim 5$, we have a widely spaced discrete spectrum (from
the point of view of TeV physics) close to the one of the
Randall-Sundrum case with cross section at a KK resonances of the form
$\sigma_{res} \sim s^3/m_ n^8$. To be definite we will take $k=10^{17}{\rm GeV}$, 
$w =e^{-35}$. In the limit of very low $x$ ($x\ltsim
0.1$) the mass of the first mode is above current experimental energies
($m_1 \sim 200 {\rm GeV}$) and the phenomenology is that discussed in 
\cite{Hew}. 
\begin{figure}[ht]
\begin{center}
\mbox{\epsfig{file=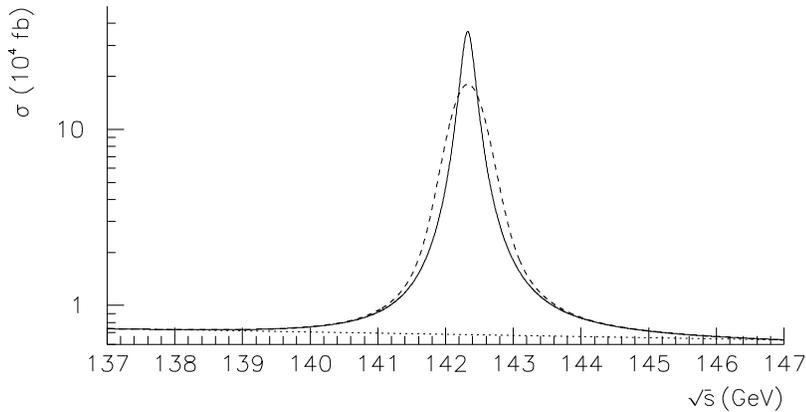,width=12cm}}
\caption{$e^+e^-\rightarrow\mu^+\mu^-$ Resonance of the third mode for 
$k=10^{17} {\rm GeV}$,
$w=e^{-35}$ and $x=1$. The
higher curve is the theoretical curve, the smaller one is the result
of the smoothing by the convolution of the theoretical curve and the
experimental resolution of the beam. The SM background lies below them.
\label{discrete}}
\end{center}
\end{figure}
For the rest of the discrete 
spectrum region ($0.1 \ltsim x \ltsim 5$)
however there are always KK resonances in the range of energies of
collider experiments. Due to the uncertainty in the collision energy the narrow 
KK peak will be reduced as shown in Fig.~\ref{discrete}. In this case the contribution of 
the KK modes gives an
excess of $\sim 10 \%$
over the SM contribution which would have been seen either by direct
scanning if the resonance is near the energy at which the experiments
actually run or by means of the process
$e^+e^-\rightarrow\gamma\mu^+\mu^-$ which scans a continuum of energies 
below the center of mass energy of the experiment. 
Thus this range of $k$, $w$, and $x$ is already excluded. Of course if $k$ 
is raised the KK modes become heavier and there will be a value for which the 
lightest KK mode is above the experimental limits. However as $k$ increases it 
reintroduces the hierarchy problem which the warp factor is designed to 
eliminate.

For values of x greater than $x\sim 5$ the spacing in the spectrum is
so small that we can safely consider it to be continuous. In this case
we substitute in $D(s)$ the sum for $n\geq2$ by an integral over the
mass of the KK excitations, \textit{i.e.}
\be
D(s)_{KK}\approx\frac{1/c_1^2}{s-m_1^2+i\Gamma_1 m_1}+
\frac{1}{\Delta m\; c^2} \int_{m_2}^{M_s}dm\; \frac{1}{s-m^2+i\epsilon}
\ee
where the value of the integral is $\sim i \pi/2\sqrt{s}$ with the principal
value negligible in the region of interest ($\sqrt{s} \ll M_s$) 
and we have considered constant coupling suppression $c$ 
for the modes with $n\geq 2$.
The first state is singled out because of
its different coupling. In fact at these values of $x$ the biggest
contribution comes from this state (the coupling of the rest being
very small). 
\begin{figure}[ht]
\begin{center}
\mbox{\epsfig{file=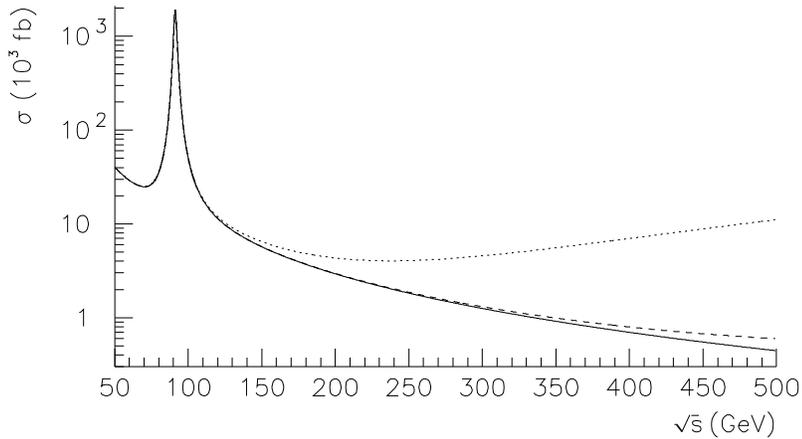,width=12cm}}
\caption{$e^+e^-\rightarrow\mu^+\mu^-$ Cross section in the continuum
limit for $k=10^{17}{\rm GeV}$ and $x=11$. The upper
curve is for $w=e^{-36}$, the medium one for $w=e^{-35}$ and the lower
one is the SM contribution (one loop calculation). The resonance we
see is the Z peak.
\label{continuum}}
\end{center}
\end{figure} 
For $w\approx e^{-35}$ the contribution
of the KK tower is negligible compared with the SM background. However
if we allow the value of $w$ to decrease, the 
coupling of the first state increases enough to give noticeable contribution
to the cross section ($\sim 30 \% \mbox{ at }\sqrt{s}=189\mbox{GeV for } 
w\approx e^{-36}$).
Thus we can exclude in this continuum limit
values of $w\approx e^{-36}$ and below. A quantitative estimate is given in
Fig.~\ref{continuum} where three curves are shown, the lowest one
corresponds to the SM alone, the second takes into account also the
tower of KK in the case of $w=e^{-35}$ and the last one is also with
the KK tower but this time with $w=e^{-36}$ which as can be seen from
the plot is already excluded.

To summarize, from the process $e^+ e^- \rightarrow \mu^+\mu^-$ we can
exclude a window of the form $0.1 \ltsim x \ltsim 5$ for the preferred
value of $w$ (this window corresponds to a value of the mass of the
first KK mode between $\sim 0.3$ GeV and $\sim 200$ GeV). Both the
the very low $x$ limit and the continuum limit
are allowed. Finally, we can also exclude values of the hierarchy
greater than $\sim e^{-36}$ except for the $0 \leq x \ltsim 0.1$ region.  

We have considered as an example the process
$e^+e^-\rightarrow\mu^+\mu^-$ but it may be possible to obtain similar or even 
stricter bounds
could be obtained from other processes like $e^+e^-\rightarrow\gamma +
\mbox{\textit{missing energy}}$ for example.  We shall present the
detailed cross-section of  the process 
$e^+e^-\rightarrow\gamma +\mbox{\textit{light KK mode}}$ in another
 publication, let us only mention here that by dimensional analysis
 the ratio of this cross-section to two-photon electron-positron 
annihilation 
\ba
\frac{\sigma(e^+e^-\rightarrow\gamma +\mbox{\textit{light KK mode}})}
{\sigma(e^+e^-\rightarrow\gamma + \gamma)} \sim
\frac{s}{c_1^2 \alpha} \sim 10^2 \frac{s}{c_1^2}
\ea
and so we see that this process may be phenomenologically very important.

A further bound on the parameters of our model can be put from the
Cavendish experiments. The fact that gravity is Newtonian at
least up to millimeter distances implies that the corrections to
gravitational law due to the presence of the KK states must be
negligible for such distances. The gravitational potential is the
Newton law plus a Yukawa potential due to the exchange of the KK
massive particles (in the Newtonian limit):
\be
V(r)=-\frac{1}{M_{\rm Pl}^2}\frac{M_1 
M_2}{r}\left(1+\sum_{n>0}\left(\frac{M_{\rm Pl}}{c_n}\right)^2e^{-m_nr}\right)
\ee

The contribution to the above sum of the second and
higher modes is negligible compared with the one 
of the first KK
state, because they have larger masses and coupling
suppressions. Thus, the
condition for the corrections of the Newton law to be small for
millimeter scale distances is:
\be
x<15-\frac{1}{2}{\rm ln}\left(\frac{-{\rm ln}w}{kw}{\rm GeV}\right)
\ee

The logarithm for any reasonable choice of the parameters $k$,$w$ gives
a contribution ${\mathcal{O}}(1)$, so we can safely say that
$x\approx15$ is the maximum brane separation allowed.

\begin{figure}
\begin{center}
\begin{picture}(420,80)(0,0)
\LongArrow(0,25)(387,25)
\G2Text(130,25){0.6}{Excluded by}{Collider Experiments}
\G2Text(330,25){0.6}{Excluded by}{Cavendish Experiments}
\Text(388,20)[l]{$x$}
\Text(-2,17)[l]{$0$}
\Text(-7,35)[l]{RS}
\Vertex(0,25){2}
\Vertex(81,25){2}
\Vertex(179,25){2}
\Text(65,17)[l]{$0.1$}
\Text(182,17)[l]{$5$}
\Vertex(277.5,25){2}
\Text(263,17)[l]{$15$}
\DashLine(178.5,38)(178.5,80){2}
\Text(50,60)[l]{Discrete spectrum}
\Text(200,60)[l]{Continuum spectrum}
\label{results}
\end{picture}
\end{center}
\caption{Excluded regions for $k=10^{17}{\rm GeV}$ and  $w=e^{-
35}$}\label{exclu:reg}
\end{figure}
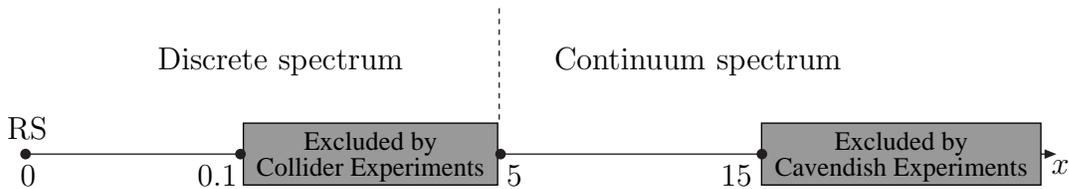

Combining the above results the allowed region of the parameters of our model
for $k=10^{17}{\rm GeV}$ and $w=e^{-35}$ are shown in Fig.~\ref{exclu:reg}.

\section{ Bi-Gravity }
Equations (\ref{m1}) and (\ref{m2}) show that, for large $x$, the lightest KK mode 
splits off from the remaining tower. This leads to an exotic possibility in 
which the lightest KK mode is the dominant source of Newtonian gravity!

Cavendish experiments and astronomical observations studying the motions of 
distant galaxies have put Newtonian
gravity to test from submillimeter distances up to distances that
correspond to $1\%$ of the size of observable 
Universe, searching for violations of the weak equivalence principle
and inverse square law. In the context of the graviton KK modes discussed above 
this constrains $m<10^{-31}{\rm eV}$ or $m>10^{-4}{\rm eV}$. Our exotic scheme corresponds 
to the choice $m_1\approx 10^{-31}{\rm eV}$ and $m_2>10^{-4}{\rm eV}$. In this case, for 
length scales less than $10^{26}{\rm cm}$ gravity is generated by the exchange of {\it 
both} the massless graviton and the first KK mode. This implies, (taking into 
account the different coupling
suppressions of the massless graviton and the first KK state) that
the gravitational coupling as we measure it is related with the
parameters of our model as:
\be
\frac{1}{M_{\rm
Pl}^2}=\frac{1}{M^2}\left(1+\frac{1}{w^2}\right)\approx
\frac{1}{(wM)^2} \Rightarrow M_{\rm Pl}\approx wM
\label{gs}
\ee

We see that the mass scale on our brane, $w M$, is now the Planck scale so, 
although the ``warp'' factor, $w$, may still be small (i.e. the fundamental 
scale  $M>>M_{Planck}$), we do not now solve even the Planck hierarchy problem. 
However our example does illustrate how gravity may be quite different from the 
form that is usually assumed. 

Using equations (\ref{gs}) and (\ref{m1}) and assuming as before that $k\approx M$, 
we find that $m_1=2ke^{-2x}\approx M_{Planck}e^{-2x}$. For $m_1=10^{-31}{\rm eV}$ we 
have $m_2\approx 10^{-2}{\rm eV}$. This comfortably satisfies the bound $m>10^{-4}{\rm eV}$ 
coming from Cavendish experiments. Here we should note that since the coupling 
of the second mode 
is 
always smaller than the one of the first mode,
the phenomenology of the continuum of
the KK states is similar to the case of Refs.\cite{D1,D2} and thus
does not conflict to
experiment.

According to this picture deviations from Newton's law will appear
in the submillimeter regime as the Yukawa corrections of the second and higher  
KK
states become important.  Also the
presence of the ultralight first KK state will give deviations
from Newton's law as we probe cosmological 
scales (of the order of the observable universe). The phenomenological signature 
of this scenario is that gravitational
interactions will appear to become weaker by the factor $w$ for distances 
larger than
$10^{26}{\rm cm}$! 

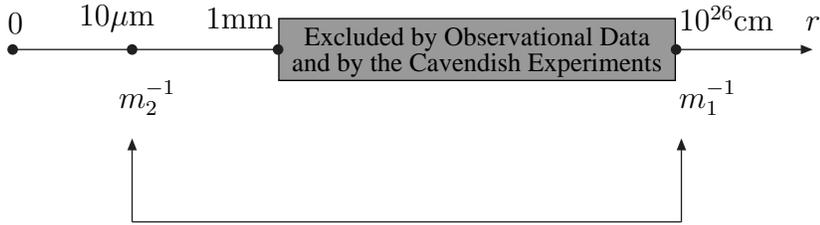
\begin{figure}
\begin{center}
\begin{picture}(320,80)(0,0)
\LongArrow(0,55)(300,55)
\G2Text(175,55){0.6}{Excluded by Observational Data}{and by the
Cavendish Experiments}
\Text(300,65)[l]{$r$}
\Text(-2,65)[l]{$0$}
\Vertex(0,55){2}
\Vertex(45,55){2}
\Vertex(100,55){2}
\Text(25,67)[l]{$10 {\rm \mu m}$}
\Text(40,37)[l]{$m_2^{-1}$}
\Text(73,67)[l]{$1 {\rm mm}$}
\Vertex(250,55){2}
\Text(252,67)[l]{$10^{26} {\rm cm}$}
\Text(252,37)[l]{$m_1^{-1}$}
\LongArrow(45,-10)(45,20)
\LongArrow(252,-10)(252,20)
\Curve{(45,-10)(252,-10)}
\label{Bi-Gravityresults}
\end{picture}
\end{center}
\caption{Exclusion regions for the Bi-Gravity case and correlation of
the first two KK states}\label{Bi-Gravity}
\end{figure}

\section{Conclusions}

In this paper we discussed a model of a  three 3-brane universe with
two positive and one intermediate negative tension brane. Our world
is confined to a positive tension brane which makes this construction
the minimal realistic model of the RS class. Due to the presence of two positive 
branes there are now two ``bound'' states, one associated with the graviton and 
one with the first KK mode. Compared to the remaining tower of KK states, the 
latter is has
relatively small mass and large coupling. Bounds on the parameter space of this 
model were placed by
comparison with data from collider and Cavendish experiments. However, the 
phenomenology of the model needs further
investigation and it may be that missing energy processes can put stricter bounds on
the parameter space.

We also explored the 
possibility of
``Bi-Gravity'' in which observable gravity is due to the exchange of both
the ordinary graviton and the first ultralight KK state.  The novel feature of 
this description is that gravity is modified at both large and small scales. In particular at large scales the strength of the gravitational force will be reduced by the warp factor. It 
is clearly of interest to explore the cosmological consequences of such a 
scheme.

{\bf Acknowledgments:} S.M., A.P. and J.S. would like to thank Peter
B. Renton and Peter Richardson for useful discussions. All authors
would like to thank Pavel Kogan for the first three figures and Subir Sarkar for 
informative discussions. J.S. would like
to thank the Department of Theoretical Physics, Oxford University, for
a kind hospitality during the course of this project. S.M.'s work is
supported by the Hellenic State Scholarship Foundation (IKY) \mbox{No. 
8117781027}.
A.P.'s work is supported by the Hellenic State Scholarship Foundation
(IKY) \mbox{No. 8017711802}. J.S.'s work is supported by M.E.C., CICYT
(Spain, AEN 96-1672) and Junta de Andaluc\'{i}a (FQM 101). The work of 
I.I.K. and G.G.R  is supported in part by PPARC rolling grant
PPA/G/O/1998/00567, the EC TMR grant FMRX-CT-96-0090 and  by the INTAS
grant RFBR - 950567.

\end{document}